\documentstyle[psfig,12pt,blois]{article}
\begin{document}
\newcommand{\ltapprox}{\raisebox{-0.5ex}{$\,\stackrel{<}{\scriptstyle\sim}\,$}}
\newcommand{\gtapprox}{\raisebox{-0.5ex}{$\,\stackrel{>}{\scriptstyle\sim}\,$}}

\heading{DETECTION AND STUDY OF HIGH-REDSHIFT \\ 
         GALAXIES THROUGH CLUSTER-LENSES }

\author{R. Pell\'o $^{1}$, J.P. Kneib $^{1}$, G. Bruzual $^{2}$, J.M. Miralles $^{3}$} 
{$^{1}$ Laboratoire d'Astrophysique de 
Toulouse, Observatoire Midi-Pyr\'en\'ees, Toulouse, France.}
{$^{2}$ Centro de Investigaciones de Astronom\' \i a (CIDA), M\'erida, Venezuela.}
{$^{3}$ Astronomical Institute, Sendai, Japan.}

\begin{bloisabstract}
We discuss a method to build up and study a sample of
distant galaxies, with $2 \ltapprox z \ltapprox 5$, using
the gravitational amplification effect in cluster-lenses
for which the mass distribution is well known. The
candidates are selected close to the critical lines at
high-z, through photometric redshift estimates on a large
wavelength range.  With respect to other methods, this
procedure allows to reduce the selection biases in luminosity, 
thanks to the cluster magnification of $\sim 1$ magnitude,
and towards active SFRs when the near-IR domain is
included. Some new results obtained with this technique
are presented and discussed.
\end{bloisabstract}

\section{Introduction}

We present the first results on the identification and study of very distant
field galaxies in the core of cluster-lenses.
Clusters of galaxies as gravitational lenses allow to study the stellar content and 
evolutionary state of galaxies much fainter than the usual spectroscopic field surveys.
The amplification close to the critical lines is typically $\Delta m \sim $ 2 to 3 
magnitudes, but it is still $\sim 1$ magnitude at 1' from the center. Sources are selected 
according to two criteria: 1) they are close to the 
critical lines at high-z, and 2) they have a photometric redshift compatible with 
$z \ge 2$. The former implies the use of well known cluster-lenses, where the
mass distribution is highly constrained by multiple images, in order to keep the 
amplification uncertainties below $\Delta m_{lensing}  \sim $ 0.3 magnitudes. 
To identify multiple images, HST images are needed to set morphological constraints,
and also photometry on a large wavelength domain. As an example, Figure 1 shows the 
central region of the cluster A2390, as well as the location of the critical lines at 
different redshifts, according to the lens model by Kneib et al. (1998). The expected 
2D distribution of arclets (mean redshift and surface density) in the particular 
cluster A2218 can be found in B\'ezecourt et al. (1998).
Some recent dicoveries on cluster-lenses strongly encourages this
approach. Among the first examples are the star-forming source $\#384$ in A2218,
at z=2.51 (\cite{E96}), and the luminous z=2.7 galaxy behind the EMSS cluster MS1512+36
(\cite{Yee96}). More recently, 3 galaxies at z $\sim$ 4 have been found in Cl0939+47 
(\cite{Tra}), a lensed galaxy at z=4.92 in Cl1358+62 (\cite{Franx97},\cite{Franx98}),
and a red lensed galaxy at z=4.04 in A2390 (\cite{Frye}, hereafter H3).

   The aim of this program is to determine the redshift distribution, the luminosity 
function and the main characteristics of the galaxy population at $z \ge 2$. For this 
exercise, we use an 

\begin{minipage}{11.0cm}
\psfig{figure=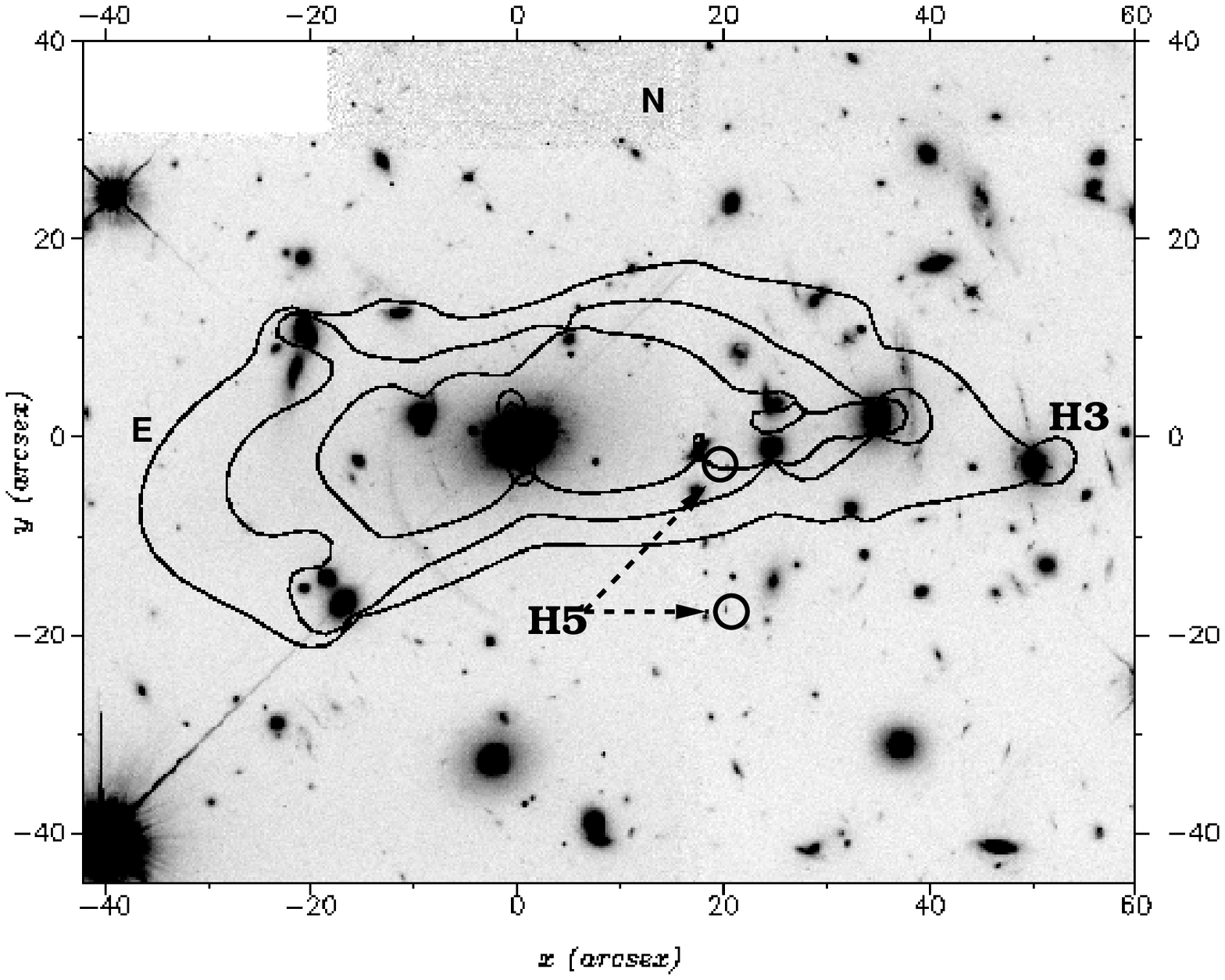,angle=270.0,height=8.0cm} 
\end{minipage}
\begin{minipage}{5.0cm}
\voffset 3.0cm
{\bf Figure 1: } HST image of A2390 (coadded $V_w$ and $I_w$ images), showing the
critical lines at z=1.0, 2.5 and 4., according to the lens model by Kneib et al. (1998).
The location of the two $z \sim 4$ galaxies H3 and H5 is also shown.
\end{minipage}

\noindent independent sample of galaxies, much less biased than the field samples 
in luminosity, or towards galaxies with a strong star-formation
activity. This is particularly true if the photometric redshift is obtained through a large
wavelength domain, including near-IR. 

\medskip

% \begin{figure}
\begin{minipage}{16.5cm}
\voffset 1.5cm
\centerline{\hbox{
\psfig{figure=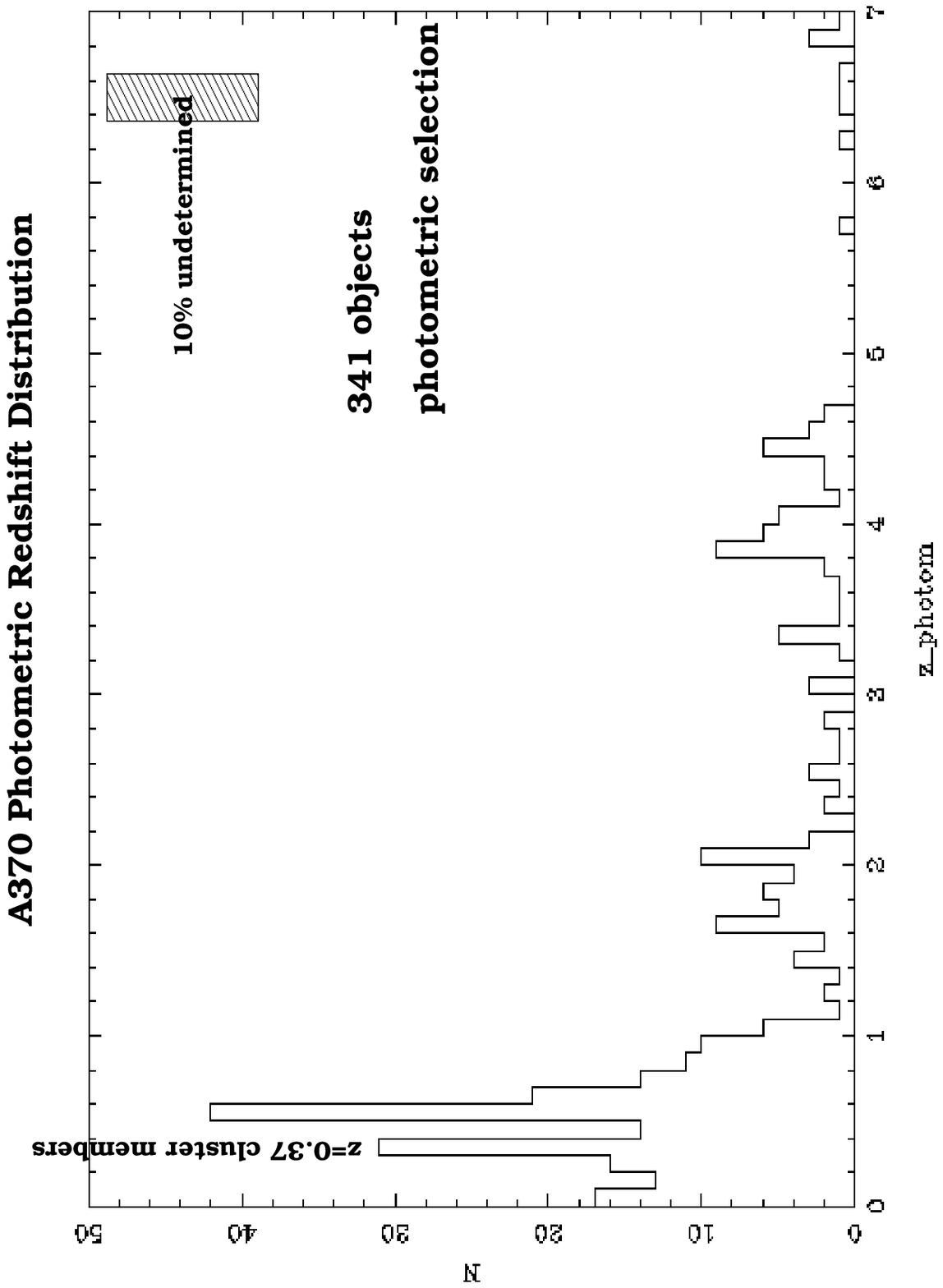,angle=270.0,height=6.0cm} 
\psfig{figure=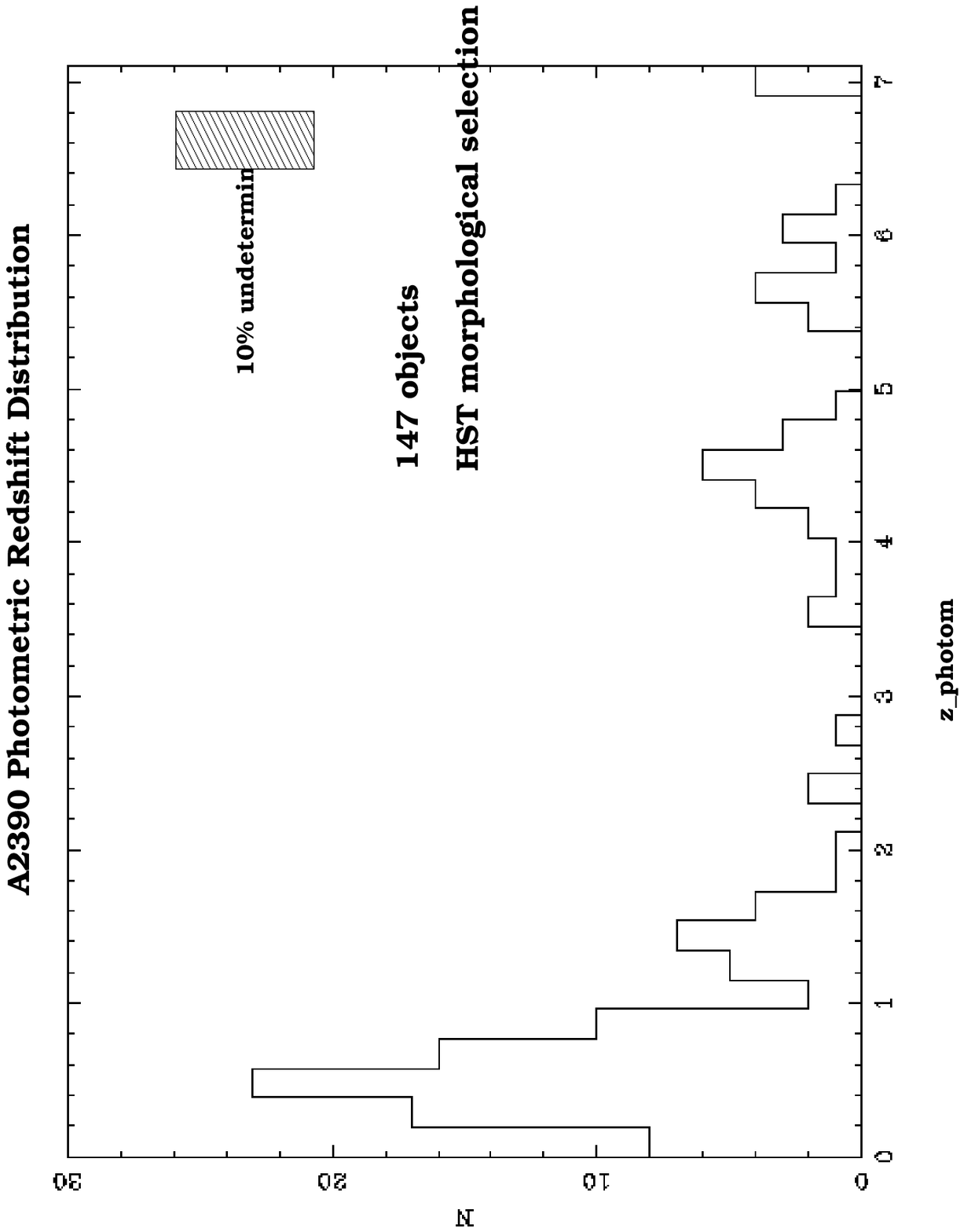,angle=270.0,height=6.0cm} 
}}
\voffset 1.5cm
{\bf Figure 2: } {\it Left:} Photometric redshift distribution in the field of A370.
{\it Right} : Photometric redshift distribution in the field of A2390 (see text).
\end{minipage}
% \end{figure}

\section{The photometric redshift approach}

Photometric redshifts were derived according to the standard minimization method 
described by Miralles \& Pell\'o (1998). The observed spectral energy distribution (SED) 
of each galaxy, as obtained from its multicolor photometry, is compared to a set of 
template spectra. The Bruzual \& Charlot evolutionary code (GISSEL98, \cite{BC}) was used to build 
5 synthetic star-formation histories, all of them with solar metallicity.
The template database includes 255 synthetic spectra. This method has allowed to identify 
sources at $z \ge 2.0$ in the field of 3 cluster-lenses: A2390, A370 and A2218. In all
cases, the photometric database includes deep near-IR J and K' images and/or U images.
As an example, we show the redshift distribution obtained in A2390 and A370 (Fig. 2). In A2390, 
the selection criteria were based on the morphology of the images
(surface brightness, elongation and orientation), and all images are located
on the very central region of the HST frames. In A370, 
the selection criteria aimed to avoid the bright 
and obvious cluster members. Nevertheless, contamination by faint galaxies at $z=0.37$
appears clearly in Fig. 2. Most of the $z \ge 2.0$ sources identified on these clusters 
are too faint to be confirmed spectroscopically using 4m telescopes.
For a selected sample of spectroscopically confirmed objects, we have tested further on the 
photometric redshift accuracy as a function of the relevant parameters, namely SFR, age and 
metallicity of the stellar population. These sources were observed during spectroscopic surveys
at CFHT, WHT and ESO (NTT, 3.6m), and they are part of an extended Toulouse/ Cambridge/ Barcelona 
collaboration. We have also cross checked on the consistency between the photometric, the
spectroscopic and the lensing redshift obtained from inversion methods (\cite{E98}). The agreement
between the three methods is fairly good up to $z \ltapprox 1.5$. For higher redshifts, a
spectroscopic sample is urgently needed to conclude.

\section{Constraining the stellar population of high-z sources}

The SEDs of these high-z sources, determined from broad-band photometry, 
can be fitted by different synthetic stellar populations. There is a degeneracy
to consider in the SFR-age-metallicity-reddening space. When the IMF and the upper 
mass limit for star-formation are fixed, the allowed parameter space can be 
constrained using the GISSEL98 code. 

\begin{minipage}{10.0cm}
\psfig{figure=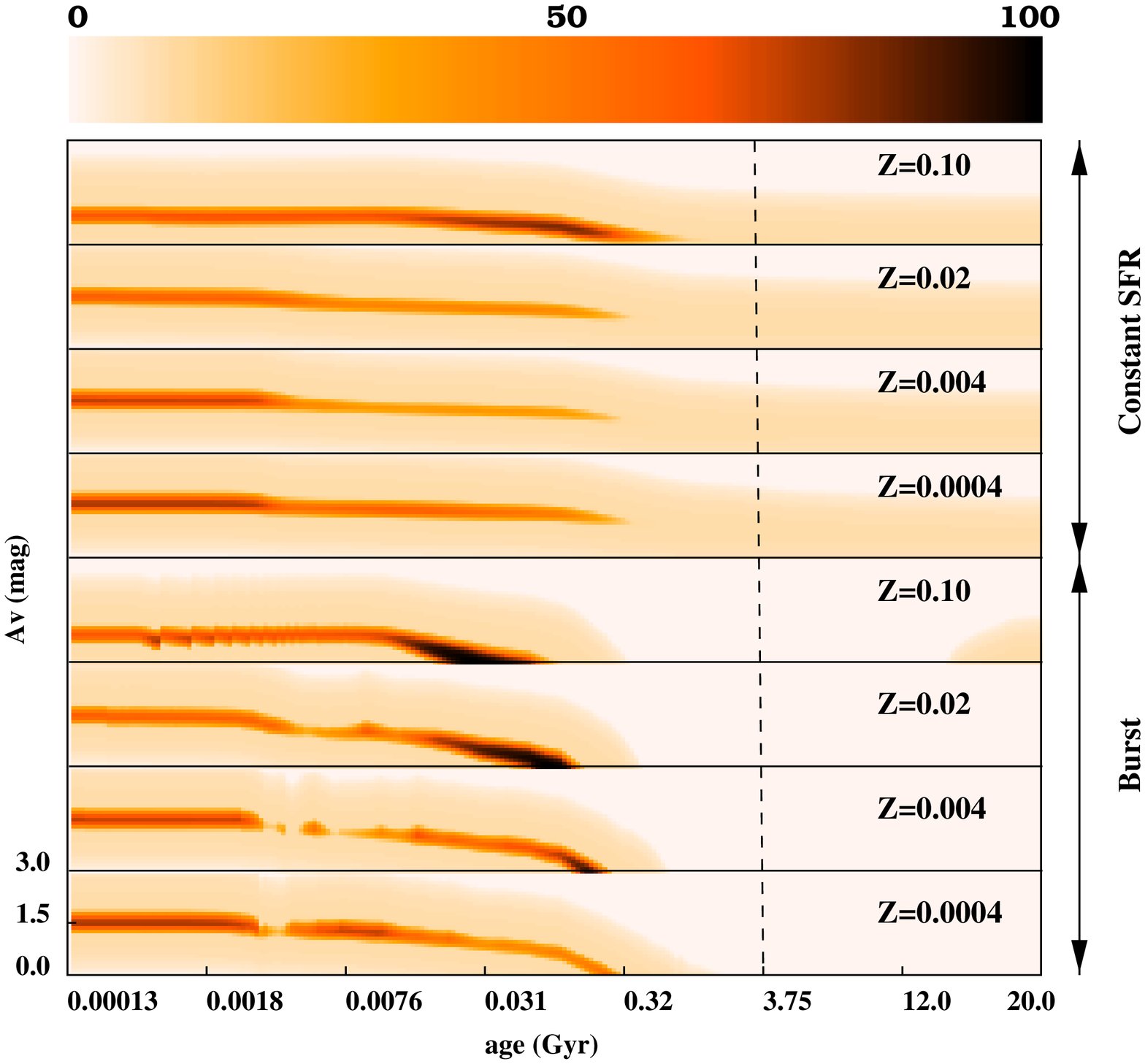,angle=270.0,height=8.0cm} 
\end{minipage}
\begin{minipage}{3.0cm}
\end{minipage}
\begin{minipage}{5.0cm}
\medskip
{\bf Figure 3: } H5 likelihood map showing in dark the most probable regions
and the degeneracy in the parameter space defined by SFR, age, metallicity
and reddening. The dotted line is the age limit corresponding to $z=4.05$
($H_0 = 50 km$ $s^{-1} Mpc^{-1}$, $q_0=0$)
\end{minipage}

We present as an example
two $z \sim 4$ sources found in A2390 (H3 and H5 in Fig. 1). The two sources are 
at the same redshift, but their stellar contents are different. The redshift of H3 was already
measured by Frye \& Broadhurst (1998), and H5 is coming from our CFHT/WHT program
(\cite{PE98}). Figure 3 displays the likelihood map of H5. H3 and H5 are
necessarily dominated by massive OB stars at the wavelengths seen in the visible
bands. Also, the presence of Ly$\alpha$ in emission points towards star-forming systems.
Two kinds of SFRs were considered: an instantaneous burst,
and a continuous star-forming system, both with the Salpeter IMF (1955), mass-limits for
star-formation $0.1M_{\odot} \le m \le 125 M_{\odot}$,
and an extinction law of SMC type (\cite{Pre}).
H3 is better constrained than H5, the weight of the old stellar population
being more important in H3 than in H5. As a consequence, it is more difficult to
reproduce the SED of H3 through an arbitrary set of the reddening value. In a
short-burst model, H3 should be older than H5. H3 and H5 are well fitted 
by burst models of 0.2 solar or even higher metallicities, but H5 is also in fair agreement 
with a reddened system undergoing a continuous star-formation activity. 
The two sources are intrinsically bright ($M^{*}_B -2$ to -1 magnitudes, depending on $A_v$),
slightly brighter than the $z \sim 4$ objects found by Trager et al. (1997) in Cl0939+4713, 
and also brighter than the $z=5.34$ galaxy of Dey et al. (1998). This fact, 
combined with the high gravitational amplification, has allowed us to obtain a spectroscopic 
redshift for them using a 4m telescope. A subsequent spectroscopic
survey using a 10m telescope will go further on this study, especially to
estimate more precisely the metallicities using the UV absorption lines.

\section{Discussion and perspectives}

The uncertainty in the amplification factor is typically 0.3 magnitudes for a large 
magnification case. Thus,
the intrinsic luminosities and stellar masses are known with an accuracy of $\sim 30\%$.
This source of error has the same importance than the model uncertainties for a relatively
well constrained SED. This gives an idea of the limitation arising 
from lens modelling when using clusters as
gravitational telescopes to access the background sources. Only the well constrained
clusters are useful for this project.

H3 and H5 are the first spectroscopically confirmed images of sources at $z \ge2$ in
A2390. The selection of high-z candidates using a photometric redshift
approach, including the near-IR bands, is strongly supported by the present results.
For most statistical purposes, photometric redshifts should be accurate enough to
discuss the properties of these extremely distant galaxies. Inversely, the spectroscopic
confirmation of the redshifts of such gravitationnally amplified sources could help
on the calibration and impovement of the photometric redshifts techniques.
A large VLT Program is presently going on, involving different european institutions
(ESO/IAP-DEMIRN/Toulouse/Roma/MPE- Heidelberg among others), and also the U. of
Chile, aimimg to perform the spectroscopic follow up of high-z 
candidates selected from the preparatory visible and near-IR photometry at the NTT and 3.6m 
telescope of ESO. 

\acknowledgements{We are grateful to J.F. Le Borgne, J. B\'ezecourt, B. Fort, 
Y. Mellier, R. Ellis, L. Campusano, G. Mathez, G. Soucail, I. Smail and B. Sanahuja for 
many discussions on this program. Part of this work was supported by the CNRS and the French PNC.}

\vfill
\end{document}